
%
%
%
%
\input harvmac
%
%
%
%
\ifx\answ\bigans
\else
\output={
  \almostshipout{\leftline{\vbox{\pagebody\makefootline}}}\advancepageno
}
\fi
%
%
%

%
%

%
%
\def\UCSD#1#2{\noindent#1\hfill #2%
\bigskip\supereject\global\hsize=\hsbody%
\footline={\hss\tenrm\folio\hss}}
%
%
\def\abstract#1{\centerline{\bf Abstract}\nobreak\medskip\nobreak\par #1}
%
%
%
%
\edef\tfontsize{ scaled\magstep3}
 \tfontsize  \tfontsize
 \tfontsize \font\titlei=cmmi10 \tfontsize
\font\titleis=cmmi7 \tfontsize \font\titleiss=cmmi5 \tfontsize
\font\titlesy=cmsy10 \tfontsize \font\titlesys=cmsy7 \tfontsize
\font\titlesyss=cmsy5 \tfontsize  \tfontsize
\skewchar\titlei='177 \skewchar\titleis='177 \skewchar\titleiss='177
\skewchar\titlesy='60 \skewchar\titlesys='60 \skewchar\titlesyss='60
%
%
%
%
%
\def\inv{^{\raise.15ex\hbox{${\scriptscriptstyle -}$}\kern-.05em 1}}
\def\lbar{{\lower.35ex\hbox{$\mathchar'26$}\mkern-10mu\lambda}} 

%
%
%
%
\def\slash#1{\rlap{$#1$}/} 
\def\dsl{\,\raise.15ex\hbox{/}\mkern-13.5mu D} 
\def\delsl{\raise.15ex\hbox{/}\kern-.57em\partial}
\def\Ksl{\hbox{/\kern-.6000em\rm K}}
\def\Asl{\hbox{/\kern-.6500em \rm A}}
\def\Dsl{\hbox{/\kern-.6000em\rm D}} 
\def\Qsl{\hbox{/\kern-.6000em\rm Q}}
\def\gradsl{\hbox{/\kern-.6500em$\nabla$}}
%
%
\def\lspace{\ifx\answ\bigans{}\else\qquad\fi}
\def\lbspace{\ifx\answ\bigans{}\else\hskip-.2in\fi} 
%
%
\def\boxeqn#1{\vcenter{\vbox{\hrule\hbox{\vrule\kern3pt\vbox{\kern3pt
        \hbox{${\displaystyle #1}$}\kern3pt}\kern3pt\vrule}\hrule}}}
%
%
\def\mbox#1#2{\vcenter{\hrule \hbox{\vrule height#2in
\kern#1in \vrule} \hrule}}
%
%
%
%

   \def\CL{{\cal L}}

%
%
%
%
%

%

\def\bar#1{\overline{#1}}

\def\bra#1{\left\langle #1\right|}
\def\ket#1{\left| #1\right\rangle}

\def\darr#1{\raise1.5ex\hbox{$\leftrightarrow$}\mkern-16.5mu #1}

%
%
\def\frac#1#2{{\textstyle{#1\over #2}}} 
%
%
%
%

\def\Tr{\mathop{\rm Tr}}

%
%
%
%

%
%
\def\ltap{\ \raise.3ex\hbox{$<$\kern-.75em\lower1ex\hbox{$\sim$}}\ }
\def\gtap{\ \raise.3ex\hbox{$>$\kern-.75em\lower1ex\hbox{$\sim$}}\ }
\def\gl{\ \raise.5ex\hbox{$>$}\kern-.8em\lower.5ex\hbox{$<$}\ }
\def\roughly#1{\raise.3ex\hbox{$#1$\kern-.75em\lower1ex\hbox{$\sim$}}}
%
%

\def\etal{\hbox{\it et al.}}

\def\np#1#2#3{{Nucl. Phys. } B{#1} (#2) #3}
\def\pl#1#2#3{{Phys. Lett. } {#1}B (#2) #3}
\def\prl#1#2#3{{Phys. Rev. Lett. } {#1} (#2) #3}
\def\physrev#1#2#3{{Phys. Rev. } {#1} (#2) #3}

\relax
\def\ref{${}^{\the\refno)}$\nref}
\def\nref#1{\xdef#1{\the\refno)}\writedef{#1\leftbracket#1}%
\ifnum\refno=1\immediate\openout\rfile=refs.tmp\fi
\global\advance\refno by1\chardef\wfile=\rfile\immediate
\write\rfile{\noexpand\item{#1\ }\reflabeL{#1\hskip.31in}\pctsign}\findarg}
\hsize 17truecm
\hoffset -0.3truecm
\vsize 24.94truecm
\voffset -1.25truecm
\ifx\epsfbox\notincluded\message{(NO epsf.tex, FIGURES WILL BE IGNORED)}
\def\insertfig#1#2#3{}
\else\message{(FIGURES WILL BE INCLUDED)}
\def\insertfig#1#2#3{
\midinsert\centerline{\epsffile{#3}}
{\baselineskip=0.45truecm{
\centerline{{#1}}\centerline{{#2}}
}}
\endinsert}\fi
%

%
\noblackbox

\def\asl{\hbox{/\kern-.6500em A}}
\phantom{a}
\vskip 4.85truecm
\centerline{RECENT RESULTS IN CHIRAL
PERTURBATION THEORY}
\centerline{FOR MESONS CONTAINING A SINGLE
HEAVY QUARK}
\vskip 0.55truecm
{\baselineskip=0.45truecm
\centerline{Elizabeth Jenkins}
\vskip 0.1truecm
\centerline{Department of Physics, University of California
San Diego, La Jolla, CA 92093 USA}
\centerline{and}
\centerline{CERN TH Division, CH-1211 Geneva 23, Switzerland}
\vfill}
\vskip -2.1truecm
\noindent
ABSTRACT
\vskip 0.55truecm

{\baselineskip=0.45truecm
\noindent
Chiral symmetry and heavy quark symmetry constrain
the interactions of mesons containing a single heavy
quark with low-momentum pions.  Chiral corrections
to the Isgur-Wise function for
$\bar B_s \rightarrow D_s^{(*)}$ versus
$\bar B \rightarrow D^{(*)}$ semileptonic decay,
the ratio of heavy meson decay constants
$f_{D_s}/ f_D$ or $f_{B_s}/ f_B$,
the amplitudes for $B_s - \bar B_s$ versus
$B^0 - \bar B{\,}^0$ mixing, and the
heavy meson mass splittings are calculated in chiral
perturbation theory.
\vskip 4.0truecm
\centerline{\it Invited talk Rencontres de Moriond 1993}
\vfill}
\UCSD{\vbox{
\hbox{UCSD/PTH 93-10}
\vskip-0.1truecm
\hbox{hep-ph/9305235}
}}{May 1993}
\eject

It is at present impossible to calculate the properties
of hadrons directly from QCD.  It is possible, however,
to obtain model-independent results by exploiting two
global approximate symmetries of QCD.  The first symmetry,
chiral symmetry,
becomes manifest in the limit of vanishing quark masses, $m_q
\rightarrow 0$.  This limit of QCD is applicable for the
three lightest quarks $u$, $d$, and $s$, and leads to the
well-known $SU(3)$ flavor symmetry amongst the light quarks.
The second symmetry, heavy quark symmetry, is relatively
new\ref\iswis{N. Isgur and M.B. Wise, \pl {232}
{1989}{113}\semi \pl {237}{1990}{527}  }.
This global approximate symmetry corresponds to the
limit of infinite quark masses, $m_Q \rightarrow \infty$, and
is manifest only if one constructs an effective field theory,
the Heavy Quark Effective Theory (HQET).  Heavy quark symmetry
is a $SU(2N_h)$ spin-flavor symmetry amongst
$N_h$ spin-1/2 heavy quarks with the same velocity in the HQET.
This formulation of QCD is relevant for the charm and bottom
quarks.

The heavy quark symmetry limit can be understood as follows.
In a meson or baryon containing a single heavy quark $Q$, the
light degrees of freedom of the hadron (everything in the
hadron except $Q$) is in the same {\it non-perturbative}
state independent of the heavy quark mass $m_Q$
and spin $S_Q$\ref\review{N. Isgur and M.B. Wise,
{\it Heavy Quark Symmetry}, Hadron '91, 549 }.
In other words, the heavy quark acts as a static color source,
with its spin decoupled from the rest of the degrees of freedom
of the hadron.
Thus, in the heavy quark symmetry limit, the light degrees of
freedom of the $B^{(*)}$ meson are the same as for a $D^{(*)}$
meson because these mesons are related by spin-flavor
symmetry transformations on the heavy quark.  Note that one cannot
calculate anything about the actual state of the light degrees of
freedom; one only knows that it is identical for mesons related
by heavy quark symmetry.

Both of these two global approximate symmetries have recently
been employed in the description of the low energy interactions
of hadrons containing a single heavy quark
with pions\ref\hqchil{M.B. Wise, \physrev {D45}{1992}{2188}
\semi G. Burdman and J.F. Donoghue, \pl {280}{1992}{287} \semi
T.M. Yan, H.Y. Cheng, C.Y. Cheng, G.L. Lin, Y.C. Lin and H.L. Yu,
\physrev {D46}{1992}{1148} }$^,$\ref\cho{P. Cho, \pl {285}{1992}{145}}.
To
describe the pion interactions of heavy quark mesons, one must
construct a chiral Lagrangian which respects heavy quark
symmetry.  Consider the pseudoscalar and vector mesons $P_Q$
and $P_Q^*$ containing heavy quark $Q$.  For a charm
quark, these are the $D$, $D^*$, while for a bottom quark, these
are the $B$, $B^*$.  The hyperfine mass splitting $\Delta = P_Q^* - P_Q
\sim 1/ m_Q$ is a $1/m_Q$ effect and is small.  (For the $D$ mesons,
$\Delta \sim 140$~MeV, while for the $B$ mesons, $\Delta \sim
45$~MeV.)  Thus, in the heavy quark limit, $P_Q$ and $P_Q^*$ are
degenerate and should be included in the chiral Lagrangian as
a single field\ref\bj{J.D. Bjorken, Les Rencontres
de Physique at La Thuile 1990, 583 }$^,$\ref\hgtasi{H. Georgi,
{\it Heavy Quark Effective Field
Theory,} Proceedings of the Theoretical Advanced Study
Institute 1991, ed. R.K. Ellis, C.T. Hill and J.D. Lykken,
World Scientific (1992) },
\eqn\hmatrix{
H^{(Q)}_a = { {(1+ \slash v)} \over 2 }
\left[ P_{a \mu}^{*(Q)} \gamma^\mu - P_a^{(Q)}
\gamma_5 \right]   .
}
This heavy meson field transforms as a $\bar 3$ under
$SU(3)$ light quark flavor, a $2$ under $SU(2)$ heavy
quark spin, and a $2$ under $SU(2)$ heavy quark flavor.
The lowest order chiral Lagrangian is given by
\eqn\lagv{
\CL_v = -i \Tr \bar H_v ( v \cdot D) H_v +
2 g \Tr \bar H_v H_v (S_{\ell v} \cdot A)
}
where
$A^{\mu} = \partial^{\mu} \pi / f + ...$
and the spin operator for the light degrees of freedom
$S_{\ell}^{\mu}$ is defined by
$\Tr \bar H_v H_v \gamma^\mu \gamma_5
= 2 \Tr \bar H_v H_v S_{\ell v}^{\mu} .$
Note that a coupling to the heavy quark spin
$\Tr \bar H_v \asl \gamma_5 H_v
=2 \Tr \bar H_v (S_{Q v} \cdot A) H_v$
is forbidden by heavy quark symmetry.  Thus,
the above Lagrangian depends on only
a single coupling constant $g$.  Heavy quark and chiral
symmetry imply that
the $D D^* \pi$ and $D^* D^* \pi$ (and the $B^{(*)}$)
couplings are related and are all given in terms of $g$.

The decay widths for the decay modes
$D^* \rightarrow D \pi$, $D^* \rightarrow D \gamma$,
and $B^* \rightarrow B \gamma$ are calculable in terms of
the coupling $g$.  Because $\Delta$ is small,
these are the dominant decay modes of the $D^*$ and $B^*$.
Note that because $B^* \slash{\rightarrow}\ \ B \pi$
(since $m_\pi > \Delta$), the $B^*B\pi$ coupling cannot
be measured directly.  The coupling is given in terms of $g$,
however, since
the $B^*B\pi$ coupling equals the $D^*D\pi$ coupling by
heavy quark symmetry.  The coupling $g$ is constrained by
experimental data\ref\accmor{The ACCMOR
Collaboration, (S. Barlag \etal), \pl {278}{1992}
{480}}$^,$\ref\cleo{The CLEO Collaboration, (F. Butler
\etal), \prl {69}{1992}{2041}}
on $D^* \rightarrow D \pi$ and $D^* \rightarrow D \gamma$,
$g^2 \ltap
0.5$\ref\amun{J.F. Amundson, C.G. Boyd, E. Jenkins, M. Luke,
A.V. Manohar, J.L. Rosner, M.J. Savage and M.B. Wise,
\pl {296}{1992}{415} \semi P. Cho and H. Georgi, \pl {296}{1992}
{408} \semi H.Y. Cheng, C.Y. Cheng, G.L. Lin, Y.C. Lin, T.M. Yan
and H.L. Yu, \physrev {D47}{1993}{1030}}.
This bound is significantly lower than quark
model predictions for $g$.

In the heavy quark limit,
$\bar B \rightarrow D^{(*)}$ semileptonic decay is described
by a single form factor, the Isgur-Wise function
$\xi( v \cdot v^\prime )$,
\eqn\semilep{\eqalign{
&\bra{D(v^\prime)} \bar c \gamma_\mu b \ket{\bar B(v) }
= \sqrt{ m_{D} m_{B} }\ \xi(v \cdot v^\prime)
(v + v^\prime )_\mu \cr
&\bra{D^*(v^\prime, \epsilon)} \bar c \gamma_\mu
\gamma_5 b \ket{\bar B(v) }
= \sqrt{ m_{D^*} m_{B} }\ \xi(v \cdot v^\prime)
[(1 + v \cdot v^\prime ) \epsilon^*_\mu
- (\epsilon^* \cdot v) v^\prime_\mu ] \cr
&\bra{D^*(v^\prime, \epsilon)} \bar c \gamma_\mu b
\ket{\bar B(v) }
= \sqrt{ m_{D^*} m_{B} }\ \xi(v \cdot v^\prime)
i \epsilon_{\mu \nu \alpha \beta} \epsilon^{*\nu}
v^{\prime \alpha} v^{\beta} .\cr
}}
The Isgur-Wise function $\xi( v \cdot v^\prime )$
describes the overlap of the light degrees of freedom
for mesons with velocity $v$ and $v^\prime$.  The
momentum transfer to leptons in the semileptonic decay
is given by $q^\mu = m_B v^\mu - m_D v^\prime$.  At
zero recoil where $v = v^\prime$ and the momentum transfer
squared to leptons is at a maximum, $q^2_{\rm max} =
(m_B - m_D)^2$, the overlap of the light degrees of freedom
is complete and the Isgur-Wise function is normalized,
$\xi( 1)=1$.  Because semileptonic $\bar B \rightarrow
D^{(*)}$ decay measures the matrix elements \semilep\
times $V_{cb}$, the normalization of the Isgur-Wise
function at zero recoil implies that $V_{cb}$ can be
extracted at zero recoil.  This procedure results in the
best determination of $V_{cb}$
to date\ref\neubert{M. Neubert, \pl {264}{1991}{455} }.
Thus, it is important to investigate how chiral corrections
affect the extraction of the Isgur-Wise function.

In the $SU(3)$ limit, the Isgur-Wise function is
independent of light quark flavor.  $SU(3)$-violating
pion loop corrections introduce light quark flavor
dependence\ref\js{E. Jenkins and M.J. Savage,
\pl {281}{1992}{331} }.  A one-loop calculation
(with $m_s \neq 0$)
implies that the Isgur-Wise functions for
$\bar B_s \rightarrow D_s^{(*)}$ and
$\bar B_{u,d} \rightarrow D_{u,d}^{(*)}$
semileptonic decay are related as follows,
\eqn\iwfn{
{ {\xi_s(v \cdot v^\prime)} \over
{\xi_{u,d}(v \cdot v^\prime)} } =
1 + \zeta( v \cdot v^\prime) {{M_K^2}
\over {16 \pi^2 f^2}}\ln\left( { {M_K^2}
\over {\mu^2}}\right)
}
where $f = 135$~MeV and
\eqn\zfn{
\zeta(v \cdot v^\prime) = \frac {30}{9}
g^2 [ r(v \cdot v^\prime) - 1 ]
}
\eqn\rfn{
r(v \cdot v^\prime) = {1 \over
{\sqrt{ (v \cdot v^\prime)^2 -1 }}}
\ln \left(v \cdot v^\prime +
{\sqrt{ (v \cdot v^\prime)^2 -1 }}\right) .
}
\vfill\eject
\insertfig{Fig. 1.  One-loop correction to the
Isgur-Wise function.  The circled cross denotes}
{the insertion of the weak current.  Wavefunction
diagrams are not shown.}
{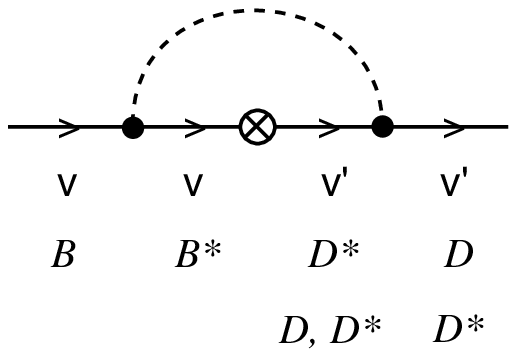}
\noindent
The diagram responsible for this correction is
depicted in \fig\iw{One-loop correction to the
Isgur-Wise function.  The circled cross denotes the
insertion of the weak current.  Wavefunction diagrams
are not shown.}.
Note that the chiral correction depends on
$(v \cdot v^\prime)$.  Because the heavy meson velocity
changes due to the weak current, the full velocity-dependent
field formulation\ref\georgi{H. Georgi, \pl{240}{1990}{447} }
of the heavy meson chiral Lagrangian
is needed for this computation.  A non-relativistic
formulation of the heavy meson chiral Lagrangian does not
suffice.  At zero recoil, $\zeta(1) = 0$ and the chiral
correction vanishes.  Thus, the normalization of $\xi(1)$
is preserved as required by heavy quark symmetry.
The chiral correction \iwfn\ is plotted as a function of
$(v \cdot v^\prime)$ in \fig\fn{Light quark dependence of
the Isgur-Wise function.} for $g^2 = 0.35$ and $\mu = 1$~GeV.
\insertfig{Fig. 2.  Light quark dependence of
the Isgur-Wise function.}{}{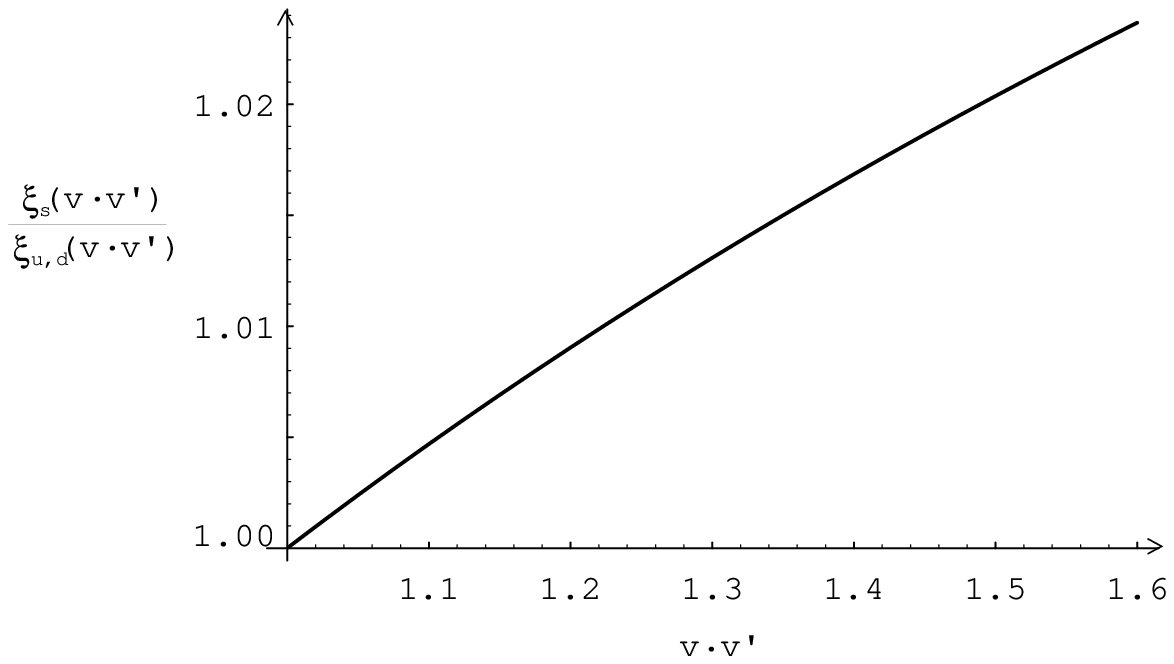}
\noindent
The correction is small;
the largest correction is $\approx 2.5\%$ at the kinematic
point $q^2=0$
or $(v \cdot v^\prime)_{\rm max} = (m_B^2 + m_D^2)/2 m_B m_D$.
The size of the chiral
correction is smaller than the anticipated size of $1/ m_c$
heavy quark symmetry-violating
corrections.
Given the present neglect of $1/ m_Q$ corrections in
the extraction of $V_{cb}$, $SU(3)$ violation also can be
neglected since it is an even smaller effect.
Isospin-breaking in the Isgur-Wise
function is expected to be even smaller than $SU(3)$ violation.

The meson decay constants $f_D$ and $f_{D_s}$ parametrize the
amplitudes for the leptonic decays $D_s^+ \rightarrow
\mu^+ \nu_\mu$ and $D^+ \rightarrow \mu^+ \nu_\mu$,
\eqn\fdef{\eqalign{
&\bra{ 0 } \bar d \gamma_{\mu} \gamma_5 c
\ket{ D^+(v) } = i f_D m_D v_\mu
\cr
&\bra{ 0 } \bar s \gamma_{\mu} \gamma_5 c
\ket{ D_s(v) } = i f_{D_s} m_{D_s} v_\mu .
\cr
}}
A chiral loop correction to the meson decay
constants\ref\fd{B. Grinstein, E. Jenkins,
A.V. Manohar,
M.J. Savage and M.B. Wise, \np {380}{1992}{369} }
can be calculated by mapping the quark current onto
the meson field current with the correct chiral and
heavy quark symmetry transformation properties,
$\bar q_a \gamma^\mu ( 1 - \gamma_5) Q
\rightarrow
{i \over 2} f_D \sqrt{m_D} \ \Tr \left( \gamma^\mu
(1 - \gamma_5) \left( H^{(Q)} \xi^\dagger \right)_a \right).
$
The diagrams for the one-loop calculation are displayed in
Ref. \fd.  For $m_s \neq 0$, the chiral correction
yields
\eqn\ff{
{ {f_{D_s}} \over {f_D}} = 1 - \frac 5 6
( 1+ 3 g^2 ) { {M_K^2}
\over {16 \pi^2 f_K^2}}\ln\left( { {M_K^2}
\over {\mu^2}}\right),
}
which implies ${ {f_{D_s}} / {f_D}} \approx 1.1$
for $g^2 = 0.4$.

The amplitudes for $B^0 - \bar B{\,}^0$ and $B_s - \bar B_s$
mixing can also be related using chiral perturbation
theory$^{\fd}$.
The mixing amplitudes are conventionally parametrized
by the meson decay constants and $B$ parameters,
\eqn\bbbar{\eqalign{
&\frac 8 3 f_B^2 m_B^2 B_B =
\bra{\bar B(v)} \bar b \gamma_\mu (1 -
\gamma_5) d  \ \bar b \gamma^\mu (1 -
\gamma_5) d \ket{B(v)}
\cr
&\frac 8 3 f_{B_s}^2 m_{B_s}^2 B_{B_s} =
\bra{\bar B_s(v)} \bar b \gamma_\mu (1 -
\gamma_5) d  \ \bar b \gamma^\mu (1 -
\gamma_5) d \ket{B_s(v)} .
\cr
}}
The four-quark operators map onto the
chiral Lagrangian operator
\eqn\fourquark{
\bar b \gamma_\mu ( 1 - \gamma_5 ) q^a
\ \bar b \gamma^\mu ( 1 - \gamma_5 ) q^a
\rightarrow \Tr \left( \left( \xi \bar H^{(b)}
\right)^a \gamma_\mu ( 1 - \gamma_5) \right)
\ \Tr \left( \left( \xi H^{(\bar b)}
\right)^a \gamma^\mu ( 1 - \gamma_5) \right).
}
Pion-loop graphs which involve just one of the
above traces correspond to renormalization of
the meson decay constant considered earlier.
Thus, it is possible to separately compute the
chiral loop renormalization of the $B$ parameter.
The calculation yields
\eqn\bb{
{ {B_{B_s}} \over {B_B} } = 1 - \frac 2 3
(1 - 3 g^2 ) {{M_K^2} \over {16 \pi^2 f_K^2} }
\ln\left( { {M_K^2} \over {\mu^2}}\right) .
}
Eq.~\ff\ and~\bb\ imply
\eqn\bff{
{ {B_{B_s} f_{B_s}^2 } \over
{B_{B} f_{B}^2 } } = 1 - ( \frac 7 3 + 3 g^2)
{{M_K^2} \over {16 \pi^2 f_K^2} }
\ln\left( { {M_K^2} \over {\mu^2}}\right),
}
which $\approx 1.3$ for $g^2 = 0.4$.

Finally, chiral corrections to the heavy meson mass splittings
can be calculated\ref\goity{J.L. Goity, \physrev {D46}{1992}
{3929}}$^{,}$\ref\rw{J.L. Rosner and M.B. Wise,
\physrev {D47}{1992}{343}}$^{,}$\ref\rs{L. Randall and E. Sather,
\pl {303}{1993}{345}}$^{,}$\ref\j{E. Jenkins,
{\it Heavy Meson Masses in Chiral Perturbation
Theory with Heavy Quark Symmetry}, to be published}.  Unfortunately, the
one-loop computation involves more unknown parameters than masses,
so there are few definite predictions.

\footatend\vfill\supereject\immediate\closeout\rfile\writestoppt
\baselineskip=0.5truecm\centerline{REFERENCES}\bigskip{\frenchspacing%
\parindent=20pt\escapechar=` \input refs.tmp\vfill\eject}\nonfrenchspacing
\bye